\documentclass[sigconf]{acmart}
\renewcommand\footnotetextcopyrightpermission[1]{}

\usepackage{amsmath,amsfonts}
\usepackage{graphicx}
\usepackage{xcolor}
\usepackage{xspace}
\usepackage{listings}
\usepackage{booktabs}
\usepackage{makecell}
\usepackage{tabularx}
\usepackage{multirow}
\usepackage{colortbl}
\usepackage{tikz}
\usepackage{nicematrix}
\usepackage{pifont}
\usepackage{float}
\usepackage{balance}
\usepackage{algorithm}
\usepackage{algpseudocode}
\usepackage{subcaption}
\usepackage{makecell}

\begin{document}

\newcommand{\name}{LLM-FSM\xspace}

\newcommand{\datasetsize}{1,000\xspace}

\newcommand{\humansize}{100\xspace}

\newcommand{\cmark}{\textcolor{green!60!black}{\ding{51}}} % ✔
\newcommand{\xmark}{\textcolor{red!90!black}{\ding{55}}}   % ✘

\newcommand{\yuheng}[1]{\textcolor{green}{\textbf{Yuheng:} #1}}

\newcommand{\berk}[1]{\textcolor{blue}{\textbf{Berk:} #1}}

\newcommand{\zhouhua}[1]{\textcolor{orange}{\textbf{Zhouhua:} #1}}

\newcommand{\peijing}[1]{\textcolor{red}{\textbf{Peijing:} #1}}

\newcommand{\caroline}[1]{\textcolor{purple}{\textbf{Caroline:} #1}}

\newcommand{\priyanka}[1]{\textcolor{green}{\textbf{Priyanka:} #1}}

\newcommand{\thierry}[1]{\textcolor{red}{\textbf{Thierry:} #1}}

\author{Xinxin Wang}
\affiliation{
  \institution{Stanford University}
  \city{Stanford}
  \state{CA}
  \country{USA}
}
\email{xxwang1@stanford.edu}

\author{Lixian Yan}
\affiliation{
  \institution{Stanford University}
  \city{Stanford}
  \state{CA}
  \country{USA}
}
\email{lxyn5869@stanford.edu}

\author{Shuhan Liu}
\affiliation{
  \institution{Stanford University}
  \city{Stanford}
  \state{CA}
  \country{USA}
}
\email{shliu98@stanford.edu}

\author{Luke Upton}
\affiliation{
  \institution{Stanford University}
  \city{Stanford}
  \state{CA}
  \country{USA}
}
\email{lupton@stanford.edu}

\author{Zhuoqi Cai}
\affiliation{
  \institution{Stanford University}
  \city{Stanford}
  \state{CA}
  \country{USA}
}
\email{zhuoqi@stanford.edu}

\author{Yiming Tan}
\affiliation{
  \institution{Stanford University}
  \city{Stanford}
  \state{CA}
  \country{USA}
}
\email{yimingt@stanford.edu}

\author{Shengman Li}
\affiliation{
  \institution{Stanford University}
  \city{Stanford}
  \state{CA}
  \country{USA}
}
\email{smli2020@stanford.edu}

\author{Koustav Jana}
\affiliation{
  \institution{Stanford University}
  \city{Stanford}
  \state{CA}
  \country{USA}
}
\email{kj2011@stanford.edu}

\author{Peijing Li}
\affiliation{
  \institution{Stanford University}
  \city{Stanford}
  \state{CA}
  \country{USA}
}
\email{peli@stanford.edu}

\author{Jesse Cirimelli-Low}
\affiliation{
  \institution{University of California, Santa Cruz}
  \city{Santa Cruz}
  \state{CA}
  \country{USA}
}
\email{jcirimel@ucsc.edu}

\author{Thierry Tambe}
\affiliation{
  \institution{Stanford University}
  \city{Stanford}
  \state{CA}
  \country{USA}
}
\email{ttambe@stanford.edu}

\author{Matthew Guthaus}
\affiliation{
  \institution{University of California, Santa Cruz}
  \city{Santa Cruz}
  \state{CA}
  \country{USA}
}
\email{mrg@ucsc.edu}

\author{H.-S. Philip Wong}
\affiliation{
  \institution{Stanford University}
  \city{Stanford}
  \state{CA}
  \country{USA}
}
\email{hspwong@stanford.edu}

\title{Heterogeneous Memory Design Exploration for AI Accelerators with a Gain Cell Memory Compiler}

\begin{abstract}

As memory increasingly dominates system cost and energy, heterogeneous on-chip memory systems that combine technologies with complementary characteristics are becoming essential. 
Gain Cell RAM (GCRAM) offers higher density, lower power, and tunable retention, expanding the design space beyond conventional SRAM. 
To this end, we create an OpenGCRAM compiler supporting both SRAM and GCRAM. 
It generates macro-level designs and layouts for commercial CMOS processes and characterizes area, delay, and power across user-defined configurations. 
The tool enables systematic identification of optimal heterogeneous memory configurations for AI tasks under specified performance metrics.

%  whose difficulty can be tuned by adjusting the state count and transition graph
% Large language models (LLMs) have shown strong potential in specification-to-RTL code generation, where a model translates a natural-language design specification into a hardware implementation. Existing benchmarks for this task are manually constructed, and current LLMs already achieve very high accuracy on them. As a result, these benchmarks can no longer effectively evaluate a model's temporal reasoning ability in hardware design or provide meaningful guidance for future training.

% In this paper, we present \name, an automated benchmark and generation pipeline for specification-to-RTL tasks. The entire process is fully automated and easily scalable by controlling the number of FSM states, enabling benchmarks that evolve with model capability. LLMs are used to generate specifications from sampled FSM graphs, while our pipeline automatically creates the corresponding RTL implementations and testbenches with full state coverage and functional correctness. 
% The final dataset includes 1,000 specification-to-RTL problems. Each example is validated through a combination of LLM-based and SAT-solver-based checks, with human verification performed on a subset. [exp details to be added.]
\end{abstract}

\maketitle

\fancyhead{}

\section{Introduction}

As SRAM scaling slows down and DRAM remains unsuitable for monolithic on-chip integration, AI accelerators increasingly face cost, area, bandwidth, and energy limits imposed by conventional memory technologies. SRAM cell sizes have largely plateaued at advanced nodes, making large arrays expensive in both area and leakage power, while off-chip DRAM -- even in high-bandwidth forms such as HBM -- cannot meet the fine-grained latency and energy demands of AI workloads. These workloads exhibit highly structured dataflows with short-lived activations and long-lived, read-mostly weights, creating a growing mismatch between application behavior and the characteristics of homogeneous memory systems. This motivates heterogeneous on-chip memory architectures that combine heterogeneous memory device technologies offering complementary density, retention, endurance, and latency characteristics, and calls for CAD frameworks capable of synthesizing memory subsystems tailored to workload needs.

Gain Cell RAM (GCRAM) is a particularly promising component of this landscape. A 2T GCRAM cell provides much higher density than 6T SRAM~\cite{complementary_gc1, complementary_gc3, liu2023gain, 16nmGC} and offers lower power with SRAM-comparable bandwidth~\cite{wwl_ls}. Its tunable retention time further broadens the design space. Si-Si GCRAM supports high-speed, short-retention use cases, while GCRAM incorporating oxide-semiconductor (OS) transistors, which have ultra-low channel leakage ($<1\times10^{-18}$ A/\textmu m), can achieve much longer retention, ranging from microseconds to hours~\cite{complementary_gc3, liu2023gain, ye2020double}. OS-based designs include hybrid OS-Si GCRAM (OS write, Si read) and OS-OS GCRAM (OS write and read). Combined with different peripheral-circuit choices (e.g., level shifters) and macro-level parameters (e.g., bank sizing), GCRAM provides rich device, circuit, and architecture knobs for tailoring area, power, speed, and retention of memory systems to diverse AI workloads.

Realizing such heterogeneous memory systems requires macro-level evaluation of area, power, speed, and retention across many device, material, and circuit configurations, which quickly exceeds what can be explored manually. To address this challenge, we introduce OpenGCRAM, an open-source compiler that automatically generates, characterizes, and implements both SRAM and GCRAM macro IPs. OpenGCRAM produces circuit designs, DRC/LVS-clean layouts, SPICE-level simulations, and detailed PPA/retention metrics for arbitrary configurations, enabling rapid and accurate exploration of heterogeneous on-chip memory compositions for AI accelerators. It also produces Verilog models, as well as .lib and .lef files for the generated memory macros, allowing seamless integration into AI accelerator design flows. The framework is fully user-modifiable and open source\footnote{We will release the code with the final version of the paper.}.

% The remainder of this paper is organized as follows: Section 2 reviews prior work on memory modeling and compiler development. Section 3 presents the standardized methodology for porting the OpenRAM compiler to new process design kits (PDKs) and memory technologies. Section 4 details the modifications made to develop OpenGCRAM and provides results on GCRAM generation and simulation. Section 5 explores future perspectives for expanding the functionality of OpenGCRAM. Finally, the paper is summarized in Section 6.

\section{Related Work}
\subsection{Heterogeneous Memory Architectures}

A substantial body of prior work demonstrates the benefits of heterogeneous memory for AI accelerators. Nguyen \textit{et al.} integrate SRAM and asymmetric 2T GCRAM within the same array to map MSBs to SRAM and LSBs to GCRAM, achieving over 48\% area reduction and 3.4$\times$ energy savings for CNN and NLP workloads~\cite{nguyen2024mcaimem}. Liu \textit{et al.} fabricated a monolithic 3D joint RRAM--gain-cell memory, where GC provides high-endurance training memory and RRAM provides low-standby-power inference memory, enabling $>$78\% standby power reduction and significant training-energy savings~\cite{RRAM_GC}. These works illustrate how mapping different data types (e.g., MSBs vs.\ LSBs, activations vs.\ weights) to technology-optimized memories can improve efficiency. However, these designs rely on handcrafted macros and lack automated tools for exploring heterogeneous memory configurations, let alone generating GDS-ready designs amenable to SPICE-level simulations and hardware validation through tapeouts.

\subsection{EDA Tools for SRAM and GCRAM}

EDA support for GCRAM remains limited. OpenRAM is the primary open-source SRAM compiler, offering netlist/layout generation and PVT characterization~\cite{OpenRAM}, but its SRAM-oriented port structure and timing model are incompatible with GCRAM's decoupled read/write operation and retention-driven behavior. Commercial solutions such as RAAAM's proprietary compiler provide 3T gain-cell macros with up to 50\% area and 10$\times$ power reduction over SRAM across 180 nm-5 nm nodes~\cite{GCRAM}, but their closed-source nature restricts research on device variants and heterogeneous-memory designs. Analytical tools like GEMTOO estimate timing, area, and bandwidth for eDRAM-like GCRAM~\cite{GEMTOO} but lack power modeling and SPICE/parasitic simulation, leading to $\approx$ 15\% error relative to post-layout results and preventing use in physical design or tapeout flows.
System-level frameworks such as GainSight~\cite{gainsight} and 3D-CIMlet~\cite{3D-CIMlet} emphasize the value of heterogeneous memories but do not synthesize or characterize GCRAM macros.

\textbf{Gap:} No existing tool provides an open, end-to-end compiler for generating, characterizing, and implementing GCRAM macros alongside SRAM. This motivates OpenGCRAM, enabling SPICE-accurate and DRC/LVS-clean exploration of heterogeneous on-chip memory architectures.

\section{Design Options for Heterogeneous On-Chip Memory Architectures}
\begin{figure}[ht]
    \centering
    \includegraphics[width=0.35\textwidth]{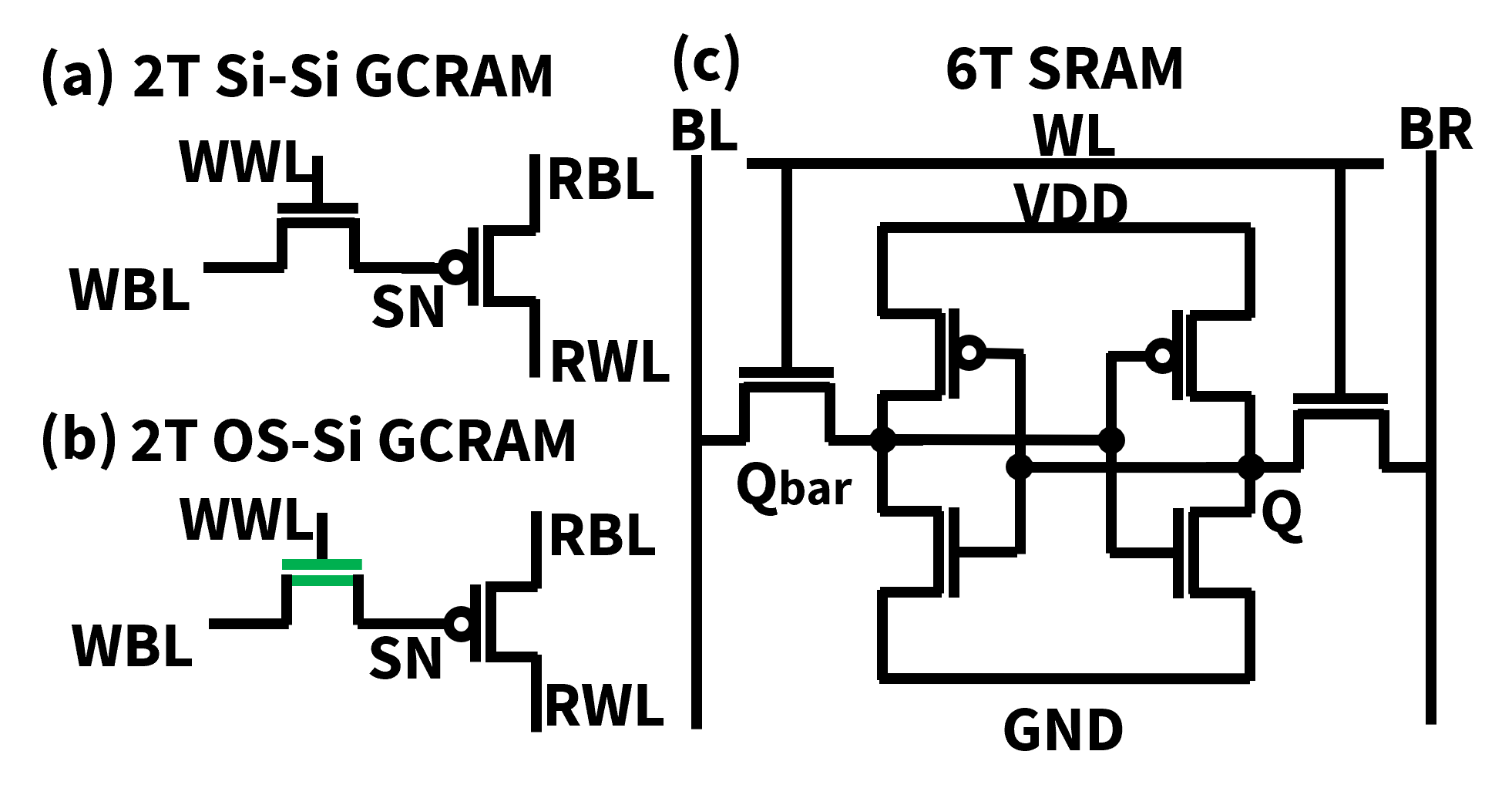} % Change "myimage.png" to your image file name
    \caption{Schematics of (a) 2T Si-Si GCRAM, (b) 2T OS-Si GCRAM, and (c) 6T SRAM bitcells. The green-colored transistor is an ultra-low-leakage OS FET.}
    \label{fig:gc_sram_schematic}
    \vspace{-1pt}
\end{figure}

\begin{figure}[ht]
    \centering
    \includegraphics[width=0.4\textwidth]{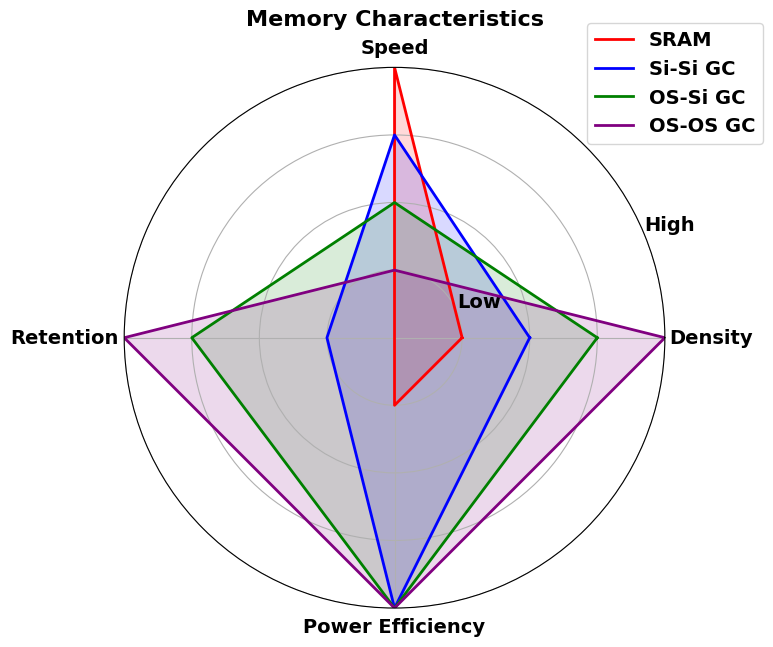} % Change "myimage.png" to your image file name
    \caption{Comparison of SRAM and GCRAM variants across density, speed, retention, and power efficiency.}
    \label{fig:memory_char}
    \vspace{-10pt}
\end{figure}

Heterogeneous memory architectures based on GCRAM and SRAM allow AI accelerators to combine technologies positioned at different trade-offs of area, speed, retention, and power. 
Figures~\ref{fig:gc_sram_schematic}(a-c) illustrate the schematics of 2T Si-Si GCRAM, 2T OS-Si GCRAM, and 6T SRAM.
A 2T GCRAM cell consists of one write transistor and one read transistor, with data stored as the voltage on the storage node (SN). During a write operation, WWL is asserted and the data value is driven onto WBL to set the SN voltage. During a read operation, RWL is toggled to sense the SN voltage, which is amplified by the gain of the read transistor and translated into a voltage change on RBL. The data retention depends on the leakage of the charge stored on SN.
The performance metrics of 2T Si-Si GCRAM, 2T OS-Si GCRAM, and 6T SRAM are summarized in Fig.~\ref{fig:memory_char}. SRAM provides the highest speed but incurs the largest area and leakage power. Si-Si GCRAM occupies a balanced point with moderate retention and power. OS-Si GCRAM achieves longer retention and lower power through an OS write device, and OS-OS GCRAM further extends retention and static-energy efficiency by using OS devices on both paths, offering the lowest leakage and highest retention at reduced access speed. These characteristics map naturally to distinct workload roles: SRAM for latency-critical operations, Si-Si GCRAM for high-bandwidth transient data, and OS-Si/OS-OS GCRAM for read-mostly or long-lived storage with stringent leakage requirements.

Peripheral circuit choices further expand the design space. Adding a level shifter (LS) improves write margins and sensing robustness, enabling higher retention and higher access speed, but increases the periphery area. Consequently, Si-Si, OS-Si~\cite{hybrid_gc}, and OS-OS~\cite{toprasertpong2023co} GCRAM each offer two design points: minimal-area no-LS variants and higher-performance LS-enabled variants. Additional circuit parameters, including sense-amplifier topology, write-driver sizing, and bank partitioning, provide further tuning of density, retention, and speed.

Together, these device and circuit options form a rich set of heterogeneous memory configurations that can be selectively composed to match the diverse dataflow requirements of modern AI accelerators.

\section{OpenGCRAM Compiler}
\label{sec:methodology}
\subsection{Compiler Implementation}
\begin{figure}[ht]
    \centering
    \includegraphics[width=0.32\textwidth]{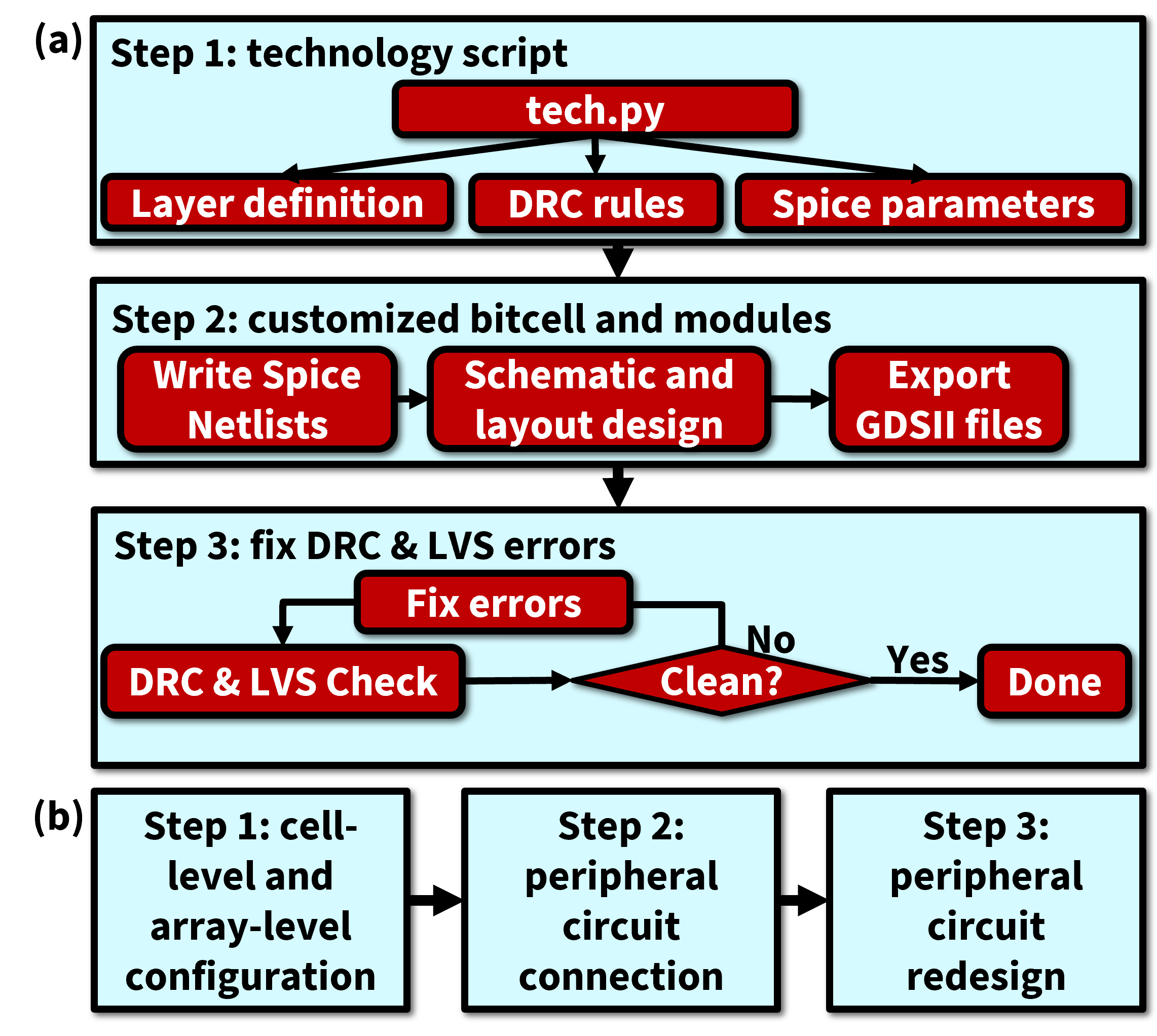} % Change "myimage.png" to your image file name
    \caption{Methodology of (a) porting the compiler to new technology nodes and (b) adding support for new memory technologies.}
    \label{fig:flow_chart}
    \vspace{-10pt}
\end{figure}

% Methodologies to get GCRAM support and new technology support capabity
As discussed previously, existing EDA tools either target SRAM only, rely on coarse analytical models, or operate at the system level. Meanwhile, heterogeneous memory architectures introduce a wide design space across device, circuit, and macro parameters. A unified tool is therefore needed to generate and validate GCRAM macros alongside SRAM and to enable systematic exploration of this space. OpenGCRAM fills this gap with an open, SPICE-accurate, DRC/LVS-clean memory-compiler framework for emerging on-chip memories.
Following the methodology in Fig.~\ref{fig:flow_chart}, we developed OpenGCRAM on top of OpenRAM. The compiler supports netlist and layout generation for 2T Si-Si GCRAM, 2T OS-Si GCRAM, and 6T SRAM in TSMC 40 nm, and provides accurate SPICE-based characterizations of delay, power, and area. It also generates Verilog models, .lib, and .lef files, enabling seamless integration with high-level synthesis and chip-design flows.

Porting OpenRAM to a new PDK requires adding a technology script that specifies device models, layer definitions, and basic design rules for SPICE netlisting, layout generation, and analytical delay estimation. Developers must also provide the SPICE and layout views for key primitives (e.g., bitcell, write driver, sense amplifier, DFF), after which macro generation becomes fully automated. DRC/LVS closure typically requires iterating on the technology script and adjusting layout functions to satisfy process-specific width, spacing, and enclosure constraints, especially at advanced nodes.

Supporting new memory types requires modifying OpenRAM's bitcell and periphery architecture to match different port configurations (e.g., differential SRAM ports, GCRAM's separate WWL/WBL and RWL/RBL, DRAM's 1T1C BL/WL, or RRAM's SL). Integration involves adding the custom bitcell views, defining port behavior, updating array-generation scripts to connect the appropriate WL/BL/SL signals, and redesigning peripheral circuits when voltage polarity, readout style, or drive requirements differ from SRAM. These modules are then incorporated into OpenRAM's address and data paths by updating module-creation and routing routines.

% \begin{figure}[ht]
%     \centering
%     \includegraphics[width=0.5\textwidth]{figures/Complementary_GC.png} % Change "myimage.png" to your image file name
%     \caption{Read waveform of the complementary NMOS-PMOS GC. The active high RWL voltage will alleviate the storage node capacitive coupling problem.(a) Schematic waveforms of write and read operations of complementary 2T GCs operating with voltage-sensing scheme.(b) Example of complementary GCRAM device structure. NMOS works as write transistor, PMOS works as read transistor.}
%     \label{fig:Complementary_GC}
% \end{figure}

\subsection{GCRAM Macro Architecture}

\begin{figure}[ht]
    \centering
    \includegraphics[width=0.4\textwidth]{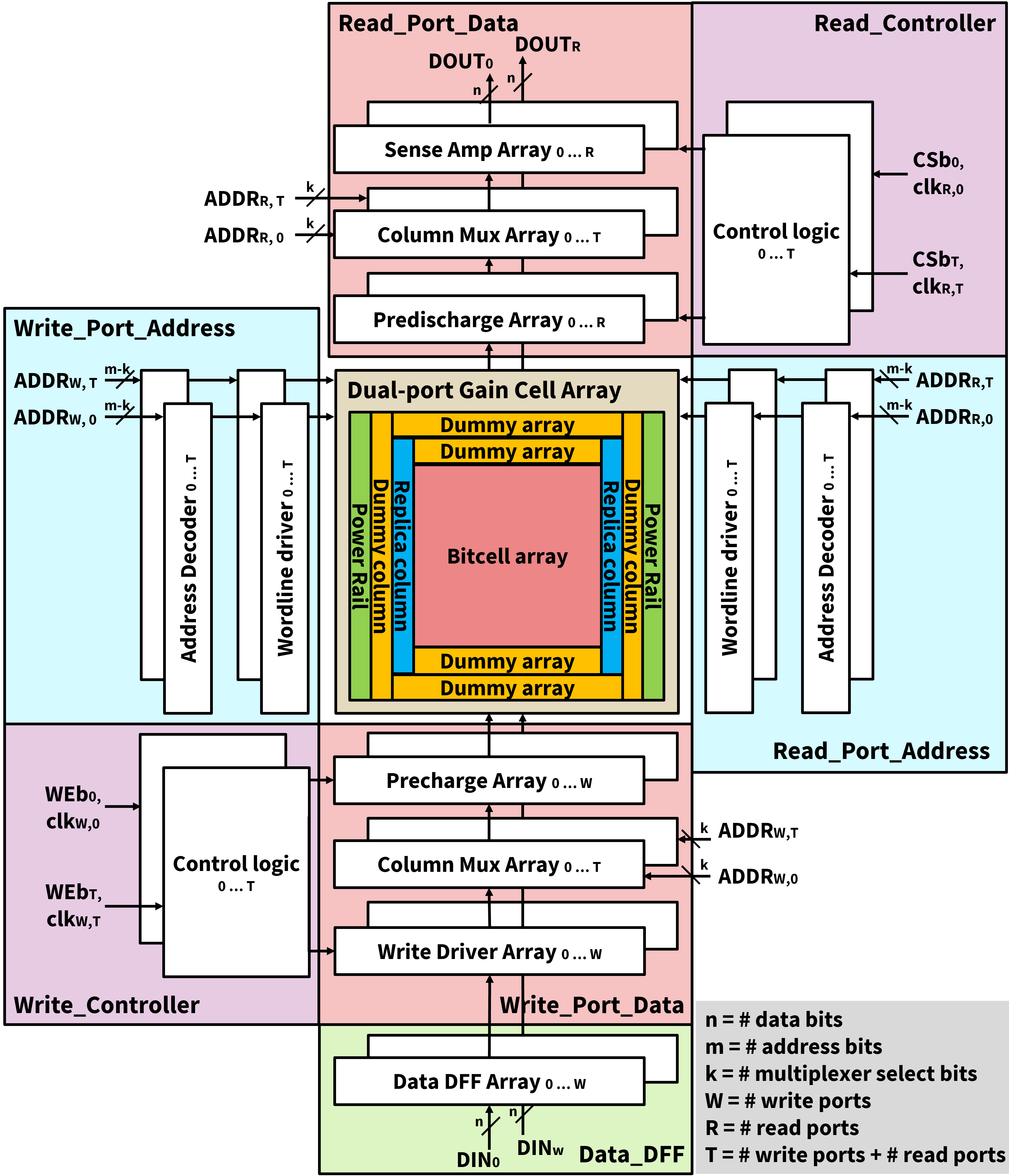} % Change "myimage.png" to your image file name
    \caption{GCRAM macro architecture adopted by OpenGCRAM.}
    \label{fig:gc_arch}
    \vspace{-5pt}
\end{figure}

The GCRAM macro architecture in OpenGCRAM uses separate read and write ports (Fig.~\ref{fig:gc_arch}). Each macro consists of a GCRAM bank, a \texttt{Data\_DFF} that captures write data, and dedicated controllers that generate read and write enable signals. Within the bank, the bitcell array is driven by \texttt{Write\_Port\_Address} and \texttt{Write\_Port\_Data} modules for WWLs and WBLs, while \texttt{Read\_Port\_Address} and \texttt{Read\_Port\_Data} modules drive the RWLs and sense the RBLs.

To support single-ended GCRAM operation, we modified OpenRAM's peripheral circuitry and BL/WL configurations. The write driver and sense amplifier are connected to a single BL; the unused $\mathrm{BL_b}$ path was removed, and the sense amplifier now references a dedicated on-chip generator~\cite{Vref}.

\begin{figure}[ht]
    \centering
    \includegraphics[width=0.4\textwidth]{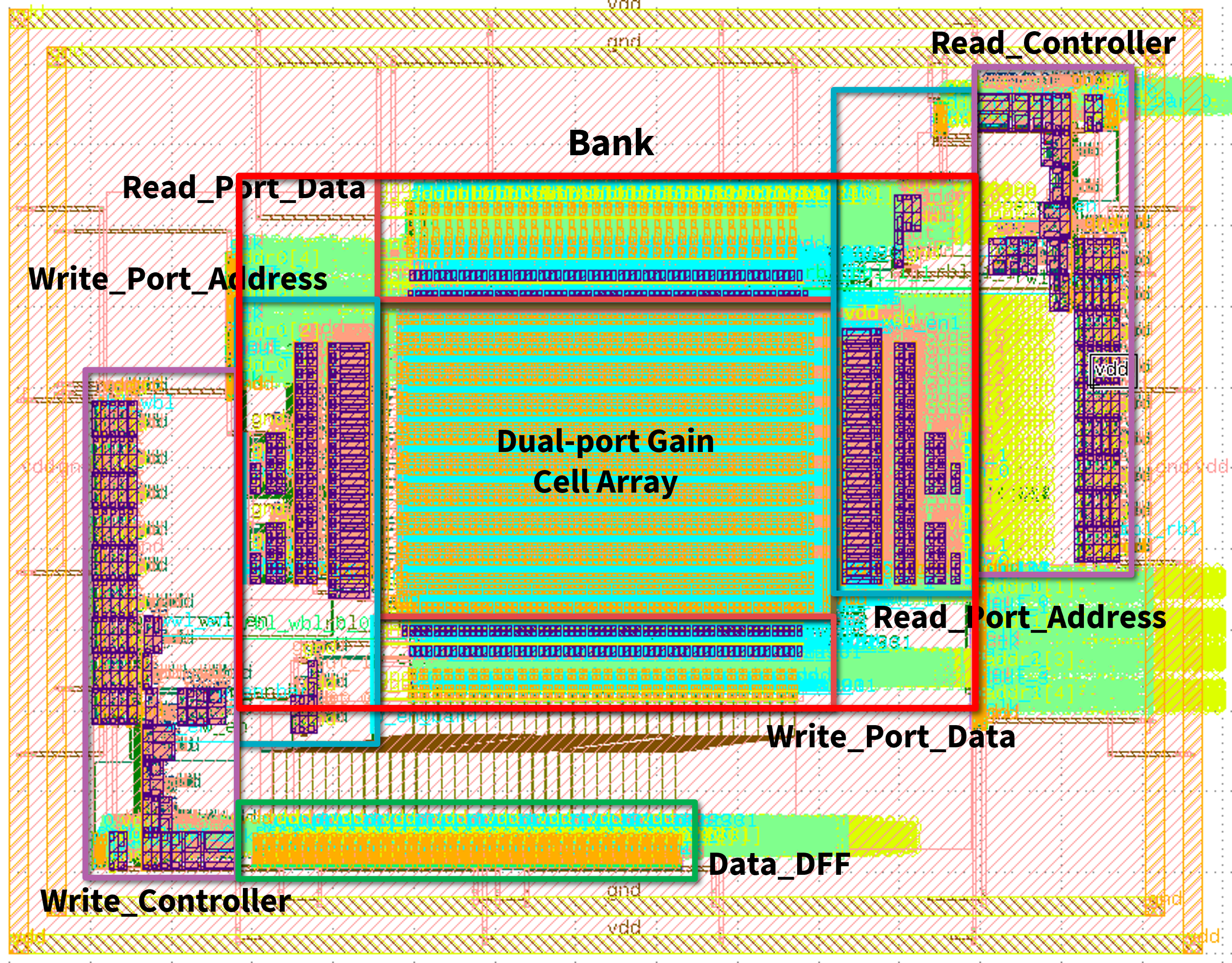} % Change "myimage.png" to your image file name
    \caption{Generated layout of a 32\texttimes 32 GCRAM macro with bitcell array, dual-port peripherals, and power rings.}
    \label{fig:layout}
\end{figure}

For GCRAM, NMOS-NMOS cells use an active-low RWL, which introduces additional SN degradation due to coupling, whereas NMOS-PMOS cells use an active-high RWL that boosts the SN voltage~\cite{complementary_gc2, complementary_gc1, hybrid_gc}. Thus, both Si-Si and OS-Si GCRAM are designed with an NMOS write transistor and a PMOS read transistor. Both GCRAM types employ a predischarge circuit that grounds the RBL prior to sensing. Since OpenRAM uses a precharge circuit for SRAM, we added scripts to generate the NMOS predischarge array and integrate it into \texttt{Read\_Port\_Data}. Because predischarge uses an active-high EN signal (in contrast to the active-low EN\textsubscript{b} for precharge), an additional inverter was inserted into the read-controller logic.
Figure~\ref{fig:layout} shows an example 32\texttimes 32 dual-port Si-Si GCRAM macro generated by OpenGCRAM with all peripheral circuits and control modules included.

\section{Design Explorations with OpenGCRAM}

\subsection{Area Comparison}

\begin{figure}[ht]
    \centering
    \includegraphics[width=0.45\textwidth]{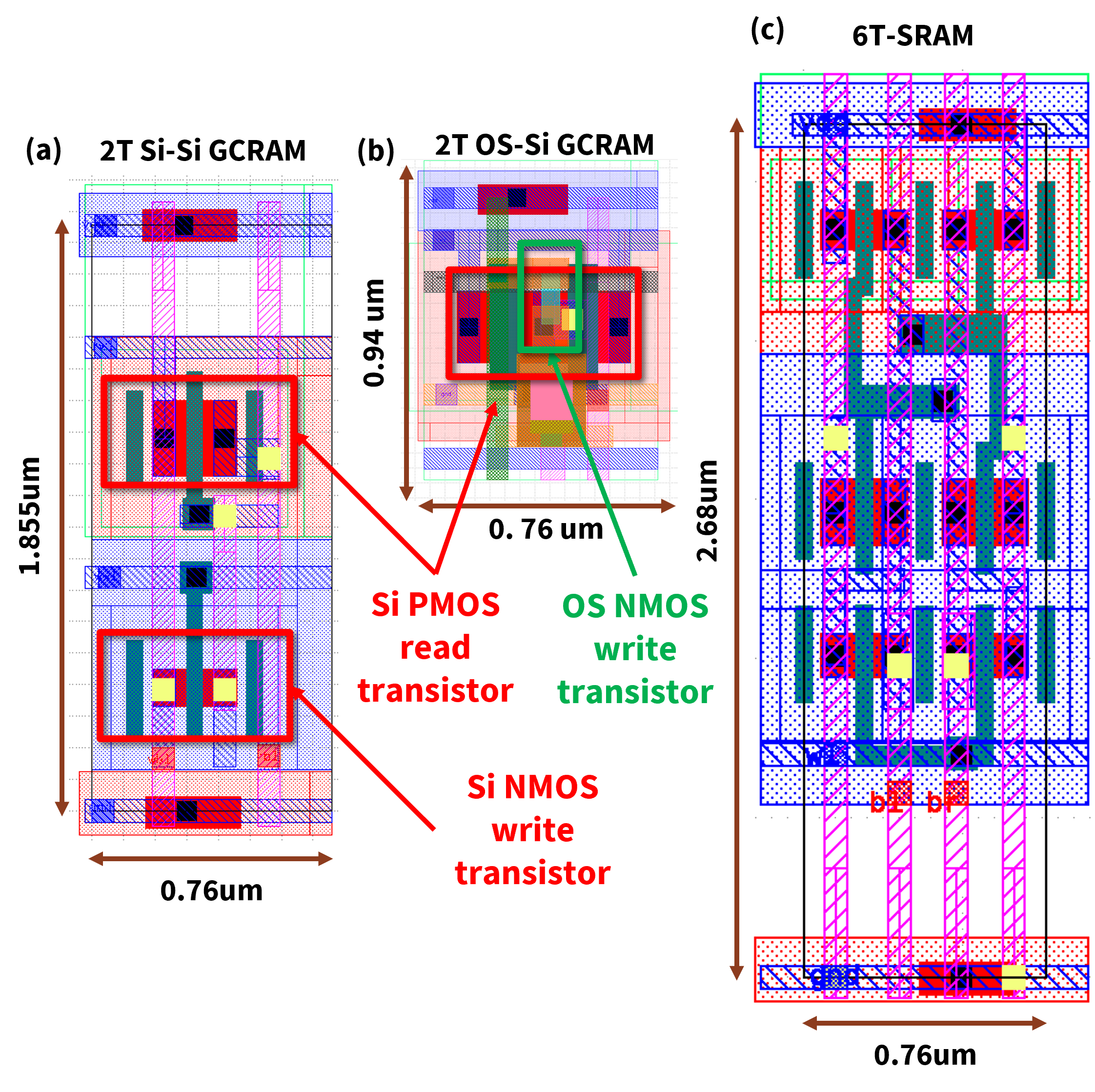} % Change "myimage.png" to your image file name
    \caption{Layouts of (a) 2T Si-Si GCRAM, (b) 2T OS-Si GCRAM, and (c) 6T SRAM cells. GCRAM cells are smaller than 6T SRAM cells.}
    \label{fig:gc_sram_layout}
    
\end{figure}

Memory capacity is one of the most important attributes for AI compute. Previous work ~\cite{hu2024co} showed OS-Si GCRAM to have 3x bit-cell density of high-density SRAM at 5-nm node. Fig.~\ref{fig:gc_sram_layout}(a-c) shows the bitcell layouts of 2T Si-Si GCRAM, 2T OS-Si GCRAM, and 6T SRAM. 2T Si-Si and 2T OS-Si GCRAM cells occupy 69\% and 35\% of the 6T SRAM cell area, respectively. Si-Si GCRAM and SRAM comply with standard logic design rules, whereas the OS write transistor in OS-Si GCRAM satisfies front-end-of-line (FEOL) constraints for tight-pitch metal-layer integration and enables monolithic 3D stacking on top of the Si read transistor using back-end-of-line (BEOL) process. 

\begin{figure}[ht]
    \centering
    \includegraphics[width=0.45\textwidth]{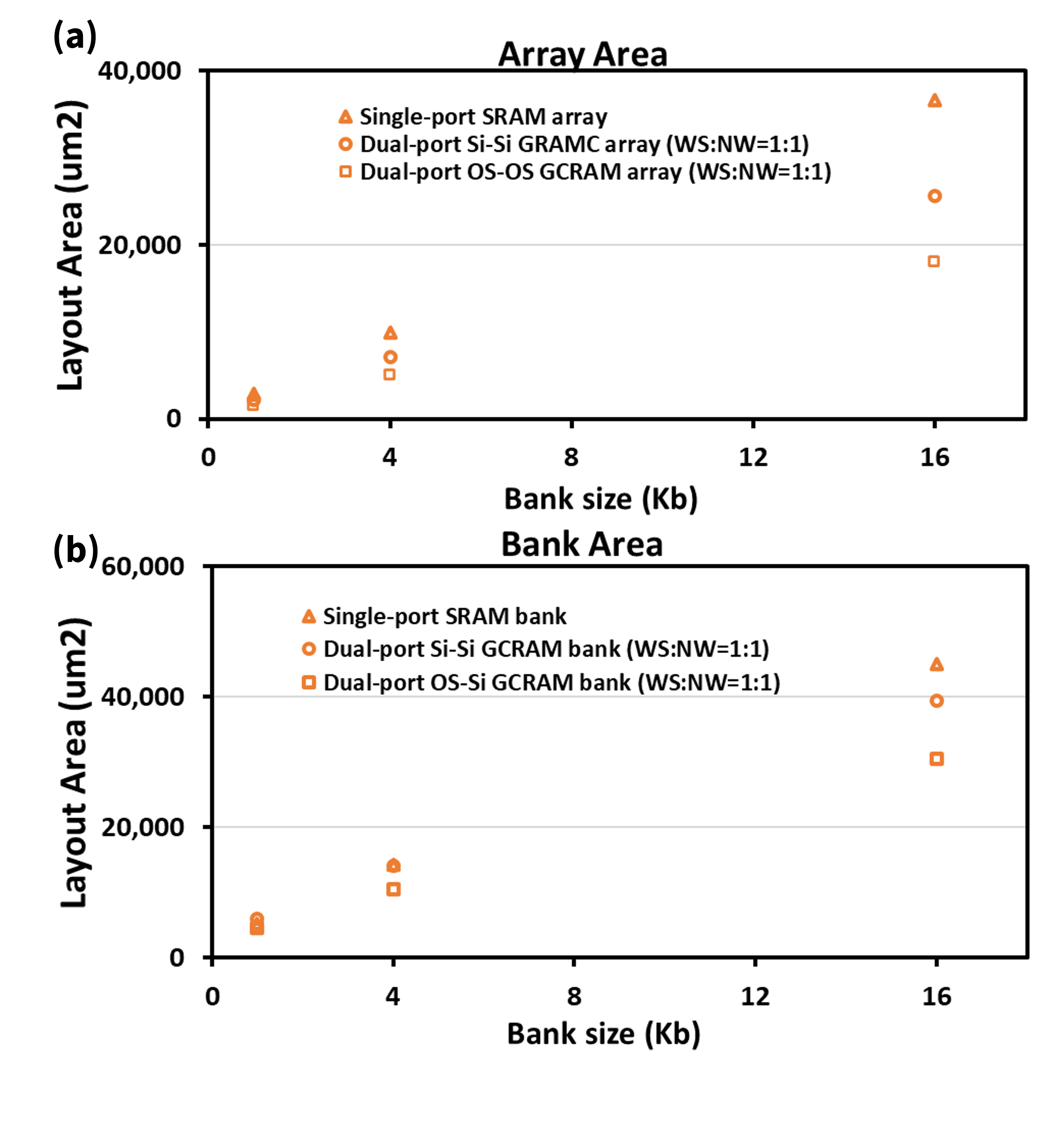} % Change "myimage.png" to your image file name
    \caption{Area comparison of dual-port GCRAM vs. single-port SRAM: (a) GCRAM arrays are smaller; (b) with peripherals, Si-Si GCRAM is smaller than SRAM above 1 Kb, while OS-Si is smallest overall.}
    \label{fig:area}
\end{figure}

Here, we present memory macro area comparison for Si-Si GCRAM, Si-Si GCRAM, and 6T SRAM cells including all peripheral circuits and control modules. We generated Si-Si GCRAM, OS-Si GCRAM and SRAM macros with multiple configurations using OpenGCRAM. Figure~\ref{fig:area}(a) compares the bitcell-array areas of dual-port GCRAM and single-port 6T SRAM (1-16 Kb). Owing to its smaller bitcell, the Si-Si GCRAM array is smaller than SRAM, and the OS-Si array is even denser. Figure~\ref{fig:area}(b) reports total bank area, including decoders, drivers, data DFFs, and sense amplifiers. Because we compare dual-port GCRAM with single-port SRAM, the GCRAM peripherals are larger; however, Si-Si GCRAM still becomes smaller than SRAM once bank size exceeds 1 Kb, and OS-Si GCRAM is the smallest across all sizes. The area advantage of GCRAM grows with bank size as peripheral overhead is amortized. In addition, GCRAM's dual-port architecture enables simultaneous read and write operations, whereas dual-port SRAM requires roughly 2\texttimes  the area of a single-port design~\cite{dual-port_sram}.

\subsection{Frequency, Bandwidth, and Power Comparison}
\begin{figure}[ht]
    \centering
    \includegraphics[width=0.42\textwidth]{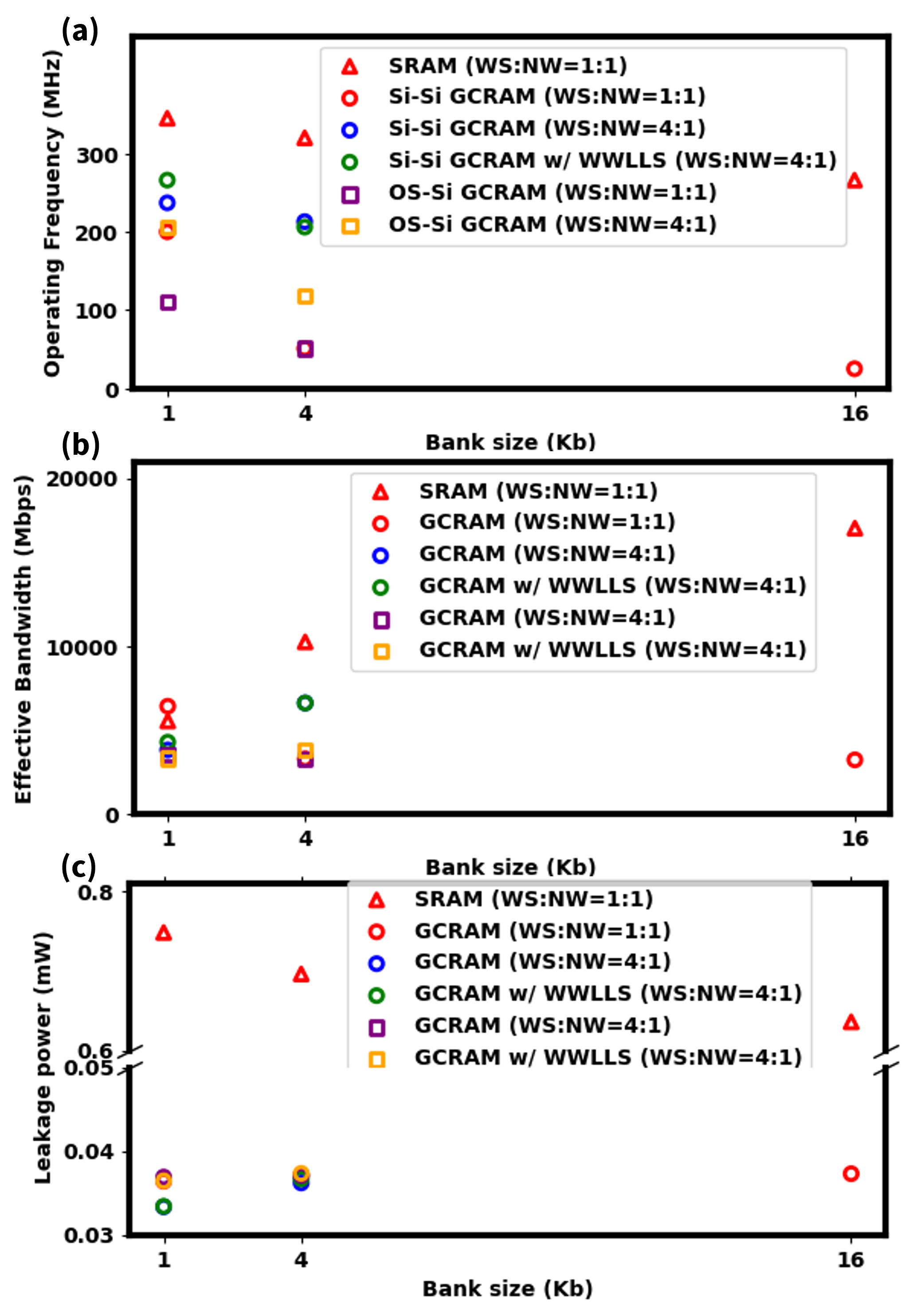} % Change "myimage.png" to your image file name
    \caption{Comparison of (a) operating frequency, (b) effective bandwidth, and (c) leakage power for SRAM, Si-Si GCRAM and OS-Si GCRAM.}
    \label{fig:freq}
\end{figure}

OpenGCRAM provides automated HSPICE simulation for accurate performance evaluation. Using compiler-generated stimuli and netlists, we benchmarked Si-Si GCRAM against 6T SRAM. As shown in Fig.~\ref{fig:freq}(a), the operating frequency of GCRAM is lower due to its single-ended read path and SN degradation by $V_T$ during write operation. The sharp frequency drop for the 1:1 configuration arises from additional delay-chain stages required for timing closure.

Macro organization also affects speed. 
When the word\_size (WZ) to num\_words (NW) ratio is 4:1, the array is naturally square and avoids column-mux overhead, resulting in a higher read frequency than the 1:1 case.
GCRAM performance can be boosted by adding a WWL level shifter (WWLLS), which raises the SN write level to VDD and improves read speed, as seen in the green points in Fig.~\ref{fig:freq}(a). The extra supply incurs an area penalty from an additional power ring (Fig.~\ref{fig:area}(a)).

Figure~\ref{fig:freq}(b) reports effective bandwidth. SRAM bandwidth is higher but reduced by the shared read/write port. GCRAM bandwidth depends on macro configuration and can be improved through peripheral-circuit optimization.

Figure~\ref{fig:freq}(c) shows that GCRAM has orders-of-magnitude lower leakage power than SRAM. This is because GCRAM bitcell has no direct VDD-GND path, resulting in negligible leakage power.

\subsection{GCRAM Retention Modulation}
\begin{figure}[ht]
    \centering
    \includegraphics[width=0.47\textwidth]{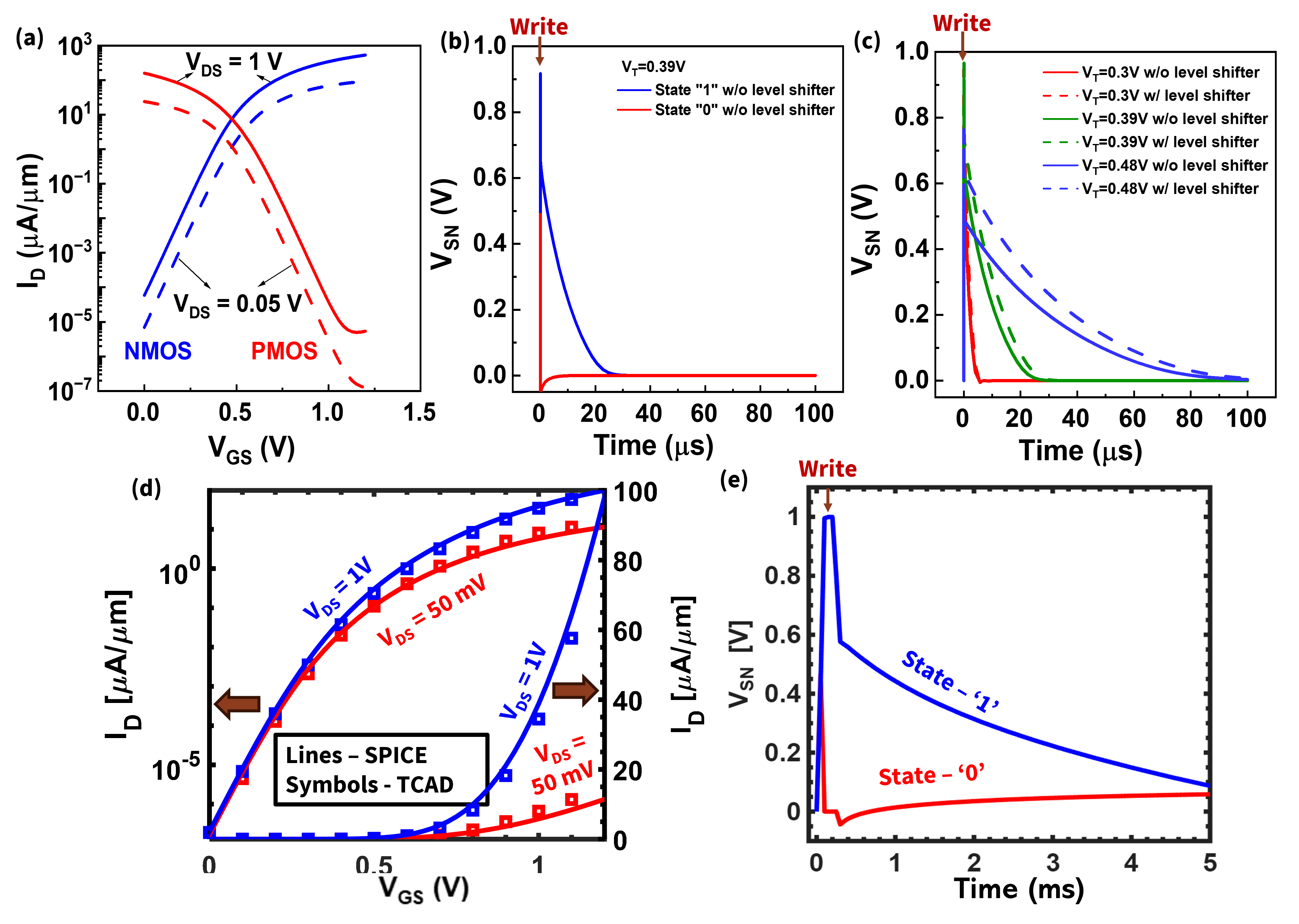} % Change "myimage.png" to your image file name
    \caption{(a) Si device $I_d$-$V_gs$. (b) Si-Si GCRAM retention. (c) Retention modulation using $V_T$ and WWLLS.
(d) OS device $I_d$-$V_gs$. (e) OS-Si GCRAM retention.}
    \label{fig:retention}
\end{figure}
GCRAM retention is dictated by the subthreshold leakage of the write transistor. Figure~\ref{fig:retention}(a) shows the Si NMOS and PMOS characteristics. Figure~\ref{fig:retention}(b) shows that Si-Si GCRAM retains data for microseconds, limited by state ``1'' decay. Retention can be tuned by increasing the write-transistor $V_T$ or by adding a WWLLS, which improves the SN write level (Fig.~\ref{fig:retention}(c)).

From a device perspective, retention depends on SN capacitance and leakage through the write and read transistors. Ultra-low-leakage OS materials (e.g., IGZO, ITO, IWO~\cite{lu2022first, complementary_gc3, liu2023gain, ye2020double}) provide orders-of-magnitude smaller off-currents ($<1\times10^{-18}$ A/\textmu m), dramatically extending retention. Using TCAD-calibrated ITO models (Fig.~\ref{fig:retention}(d)), OS-Si GCRAM achieves millisecond-level retention (Fig.~\ref{fig:retention}(e)) and can exceed 10~s with further $V_T$ engineering~\cite{liu2023gain}. This tunability makes GCRAM applicable across a range of lifetimes—from microsecond-scale activation caches to hour-scale weight storage in AI inference~\cite{yue2024wkvquant}.

% \begin{table}
%   \centering
%   \caption{AI workloads evaluated by the profiler}
%   \label{tbl:ai_workloads_id}
%   \includegraphics[width=0.7\linewidth]{figures/ai_workloads_id.png}
% \end{table}

% \subsection{Performance optimization}
% WWLLS: improve operating frequency and retention
% OS-OS GC: veiloga model, performance
\subsection{Heterogeneous Memory Design Space Exploration}
\begin{table}[t!]
    \centering
    \caption{AI workloads evaluated by the GainSight profiler}
    \label{tab:ai_workloads_id}
    \resizebox{\columnwidth}{!}{%
    \begin{tabular}{p{0.3cm}lll}
    \toprule
    \multicolumn{1}{l}{\textbf{\begin{tabular}[c]{@{}l@{}}Task \\ ID\end{tabular}}} &
    \multicolumn{1}{l}{\textbf{Task Name}} & \multicolumn{1}{l}{\textbf{Test Suite}} & \multicolumn{1}{l}{\textbf{Description}} \\ \midrule
    \texttt{1} & 2dconvolution & PolyBench~\cite{grauer2012auto,pouchet2013polyhedral} & 2D Convolution \\ \midrule
    \texttt{2} & 3dconvolution & PolyBench & 3D Convolution \\ \midrule
    \texttt{3} & llama-3.2-1b & ML Inference~\cite{reddi2020mlperf} & \begin{tabular}[c]{@{}l@{}}Meta's text-based LLM with 1 \\ billion parameters~\cite{touvron_llama_2023}\end{tabular} \\ \midrule
    \texttt{4} & llama-3.2-11b-vision & ML Inference & \begin{tabular}[c]{@{}l@{}l@{}}Meta's LLM with integrated vision \\ adapter for image recognition, \\ total of 11 billion parameters~\cite{touvron_llama_2023}\end{tabular} \\ \midrule
    \texttt{5} & resnet-18 & ML Inference & \begin{tabular}[c]{@{}l@{}}CNN for image recognition with \\ 18 layers~\cite{he_deep_2015} \end{tabular}\\ \midrule
    \texttt{6} & bert-uncased-110m & ML Inference & \begin{tabular}[c]{@{}l@{}l@{}}"Bidirectional Encoder Representation \\ for Transformers"~\cite{devlin2019bert}, text-based \\ LLM with 110 million parameters\end{tabular} \\ \midrule
    \texttt{7} & stable-diffusion-3.5b & ML Inference & \begin{tabular}[c]{@{}l@{}}Text-to-image transformer model with \\ 3.5 billion parameters~\cite{esser2024scaling} \end{tabular} \\ \bottomrule
    \end{tabular}%
    }
\end{table}

The OpenGCRAM compiler enables fast and accurate GCRAM macro generation and performance simulation, making it easy to perform design-space exploration to meet the requirements of various workloads. Using the GainSight profiling framework~\cite{gainsight}, we extracted the performance demands for L1 and L2 caches across different AI workloads. As shown in Table~\ref{tab:ai_workloads_id}, the evaluated workloads fall into two categories: (1) kernel-level microbenchmarks (Tasks 1-2), such as 2D/3D convolution, which characterize fine-grained memory access patterns, and (2) complete AI models (Tasks 3-7), which capture system-level memory behavior in realistic deployments. Including both categories allows us to examine memory design targets across multiple levels of granularity. 

\begin{figure}[ht]
    \centering
    \includegraphics[width=0.45\textwidth]{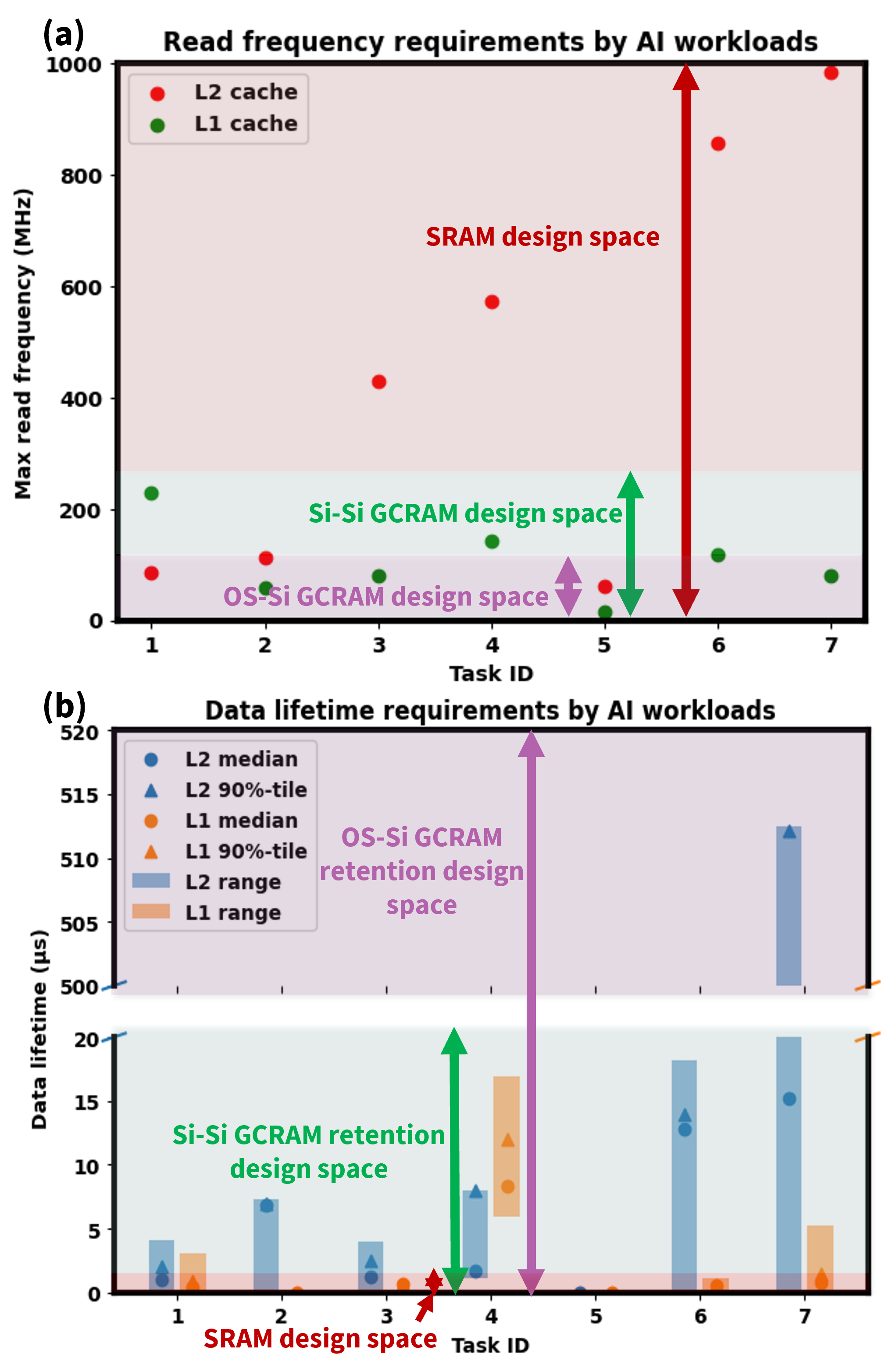} % Change "myimage.png" to your image file name
    \caption{L1/L2 cache requirements for AI workloads differ in (a) read frequency and (b) data lifetime, aligning them with the design spaces of different memory technologies.}
    \label{fig:l1_l2_cache}
\end{figure}

The maximum read frequencies and data lifetime ranges required for the L1 and L2 caches are summarized in Fig.~\ref{fig:l1_l2_cache}, with the profiling evaluated on NVIDIA H100 and scaled for NVIDIA GeForce GT 520M. Counterintuitively, most L2 tasks require much higher read frequencies than those utilizing the L1 cache. 
This is because the L2 cache is shared by all SIMD cores in the GPU, while each core has its own dedicated L1 cache. Hence, the L2 cache needs to handle many more read and write requests from multiple cores.

Figure~\ref{fig:l1_l2_cache}(a) reports the read-frequency requirements for L1 and L2 caches across all tasks, while Fig.~\ref{fig:l1_l2_cache}(b) shows the corresponding distributions of data-lifetime requirements. These performance metrics map directly onto the capability space of different memory technologies as characterized by OpenGCRAM. \textbf{Together, these results indicate that no single memory technology can efficiently cover the entire design space; instead, heterogeneous memory configurations are needed to match the diverse frequency-retention needs within each workload.}

Because higher-speed and higher-retention memory types naturally cover the design space of lower-speed ones, the selected technology should prioritize power efficiency and density. 
When multiple technologies satisfy the speed constraints, OS-Si GCRAM should be chosen first, followed by Si-Si GCRAM. 
Using these criteria, the optimal heterogeneous L1/L2 cache configuration for each task is summarized in Table~\ref{tab:cache_config}. For each task, the L1 and L2 caches should be implemented using heterogeneous memory types to meet their respective frequency and data-lifetime requirements and minimize power consumption and area.

% In your preamble:
% \usepackage{booktabs}

\begin{table}[ht]
\small
\centering
\caption{Optimal L1/L2 cache configuration by task}
\label{tab:cache_config}
\begin{tabular}{p{0.3cm}c l l}
\toprule
\textbf{Task ID} & \textbf{L1 cache} & \textbf{L2 cache} \\
\midrule
1 & Si--Si GCRAM                          & OS--Si GCRAM \\
2 & OS--Si GCRAM                          & Si--Si GCRAM \\
3 & OS--Si GCRAM                          & Si--Si GCRAM + SRAM \\
4 & Si--Si GCRAM                          & Si--Si GCRAM + SRAM \\
5 & OS--Si GCRAM                          & OS--Si GCRAM \\
6 & Si--Si GCRAM                          & Si--Si GCRAM + SRAM \\
7 & OS--Si GCRAM                          & OS--Si GCRAM + Si--Si GCRAM + SRAM \\
\bottomrule
\end{tabular}
\end{table}

\begin{figure}[ht]
    \centering
    \includegraphics[width=0.45\textwidth]{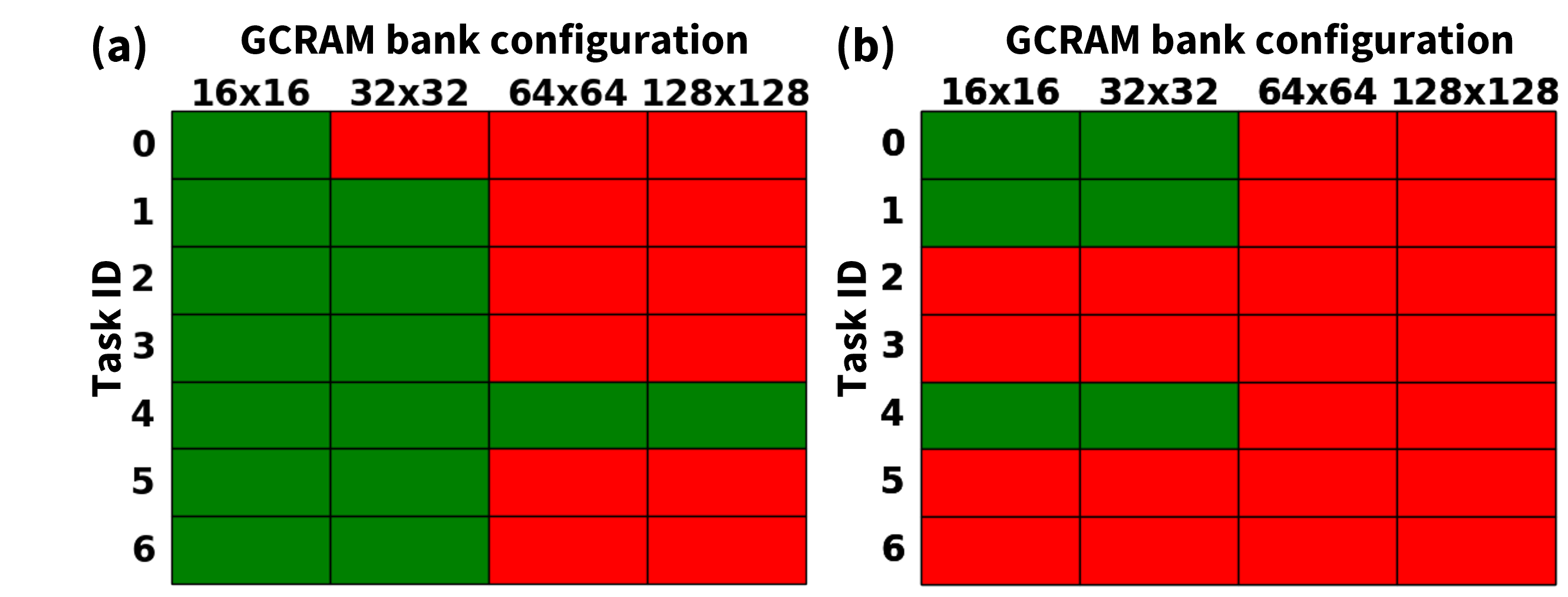} % Change "myimage.png" to your image file name
    \caption{Design choices for implementing (a) L1 and (b) L2 caches with single-bank Si-Si GCRAM. Green: workable; Red: not workable.}
    \label{fig:rshmoo}
\end{figure}

Memory macro configuration is another important hardware design parameter. Figure~\ref{fig:rshmoo} illustrates this through Shmoo plots generated using OpenGCRAM, showing the feasible Si-Si GCRAM macro configurations for each task. The x-axis represents macro size (WZ\texttimes NW), ranging from 16\texttimes 16 to 128\texttimes 128, and the y-axis corresponds to the task IDs in Table~\ref{tab:ai_workloads_id}. Each point indicates whether a given macro configuration meets the task's operating requirements (green: workable; red: not). This design-space exploration can be similarly extended to OS-Si GCRAM and SRAM using OpenGCRAM's generation and simulation capabilities.

% \subsection{Performance optimization}
% WWLLS: improve operating frequency and retention
% OS-OS GC: veiloga model, performance
% \subsection{Design Space Exploration for AI workloads}
% AI workload requirements on L1 read frequency; shmoo plot

\section{Future Work}
% The characteristics of these GCRAM varieties are illustrated in Table \ref{tab:gc_variants}.  
% \begin{figure}[ht]
%     \centering
%     \includegraphics[width=0.45\textwidth]{figures/gc_table.png} % Change "myimage.png" to your image file name
%     \caption{The comparison between different variants of 2T GCRAM .}
%     \label{fig:table}
% \end{figure}
% % change this to table later

% \begin{table}[h!]
%     \centering
    
%     \begin{tabularx}{0.475\textwidth}{|C{2.5cm}|C{1cm}|C{0.8cm}|C{1.31cm}|C{1cm}|}
%         \hline
%           & Write Speed (ns) & Read Speed (ns) & Retention (room temp) & Cell Size ($\mathrm{\mu m^2}$) \\ \hline
%         Si-Si GCRAM   & 1 & 1 & \textasciitilde 10 $\mathrm{\mu s}$ & 1.410   \\ \hline
%         OS-OS GCRAM   & 10 & 10 & \textasciitilde  10 s & 0.192   \\ \hline
%         Hybrid OS-Si GCRAM & 10 & 1 & \textasciitilde  100 ms & 0.125 \\ \hline
%     \end{tabularx}
%     \caption{Simulation results on 40 nm node 2T GCRAM variants.}
%     \label{tab:gc_variants}
% \end{table}

Beyond 2T Si-Si GCRAM, 2T OS-Si GCRAM, and 6T SRAM, heterogeneous memory architectures can be further expanded by incorporating additional technologies such as OS-OS GCRAM and RRAM. OS-OS GCRAM offers higher density and longer retention than other GCRAM variants, while RRAM naturally serves as long-term storage thanks to its non-volatile property. 
At the circuit level, alternative peripheral designs, such as voltage-mode~\cite{somasekhar20082} versus current-mode~\cite{csa} sense amplifiers, provide additional workload-specific optimization opportunities.

To broaden the design space for AI accelerators, we will integrate these device-, bitcell-, and circuit-level options into OpenGCRAM. The methodology developed for extending the compiler to new PDKs and memory technologies already supports multiple GCRAM variants and is not limited to TSMC 40 nm. It can scale to advanced nodes (e.g., TSMC 5 nm) and other memory technologies, including 1T1C DRAM and RRAM. 

% [outline of this section: 1. 2T GCRAM has many variants. Not limited to Si GC. Using oxide semiconductors extend the range of tunability, as shown in table x, by providing longer retention and more importantly more design room. 2. OpenGCRAM is a general tool that can support all these variants by .... 3. OpenGCRAM will be able to choose one of the variants based on the application spec, so that more application ]

\section{Conclusion}

As memory increasingly dominates system cost, power, and performance, heterogeneous on-chip memory architectures that combine devices with complementary characteristics are becoming essential. 
GCRAM provides higher density, lower leakage, and tunable retention, significantly broadening the design space beyond conventional SRAM. We created the OpenGCRAM memory compiler to systematically explore this design space. 
OpenGCRAM is capable of generating circuit-level designs, layouts, and macro-level area, delay, and power characterizations across both SRAM and multiple GCRAM variants. 
Through this exploration, we identified optimal heterogeneous memory types and topologies for diverse AI workloads by analyzing trade-offs in power, speed, retention, and area of various memory configurations. This heterogeneous memory design-space exploration can be further extended to additional memory technologies and peripheral-circuit options.
By integrating them into the OpenGCRAM compiler, we can pave the way toward scalable, workload-optimized on-chip memory systems.

\clearpage
\bibliographystyle{ACM-Reference-Format}
\bibliography{references}

\end{document}